\documentstyle[preprint,eqsecnum,prd,aps]{revtex}
\begin{document} 
%%%%Start of Text%%%%%%%%%%%%%%%%%%%%%%%%%%%%%%%%%%%%%%%%%%%%%%%%%%%%%%%%%%%%
%\rightline{
\preprint{
\vbox{
\halign{&##\hfil\cr
	& ANL-HEP-PR-98-39 \cr
	& JLAB-THY-98-22 \cr
	& June 2, 1998 \cr}}
}
\title{Spin Dependence of Associated Production of a Prompt Photon and a 
Charm Quark at Next-to-Leading Order in QCD}

\author{Edmond L. Berger$^a$ and Lionel E. Gordon$^{b,c}$}
\address{$^a$High Energy Physics Division, Argonne National Laboratory,
	Argonne, IL 60439\\
$^b$ Jefferson Lab, Newport News, VA 23606\\
$^c$ Hampton University, Hampton, VA 23668}
\maketitle
\begin{abstract} 
A second order, $O(\alpha ^2_s)$, calculation in perturbative quantum 
chromodynamics is presented of the longitudinal 
spin dependence of the cross section 
for the two particle inclusive reaction $p + p \rightarrow \gamma + c + X$
for large values of the transverse momentum of the prompt photon and charm 
quark.  Differential distributions are provided for the spin-averaged cross 
section and for the two-spin longitudinal polarization asymmetry $A_{LL}$ at 
the energy of the Brookhaven Relativistic Heavy Ion Collider.  An assessment 
is given of the prospects for determination of the spin dependence of the 
charm quark density.  
%\end{abstract}
\vspace{0.2in}
\end{abstract}
\narrowtext

\section{Introduction}
Because they couple in a point-like fashion to quarks,
the observation of photons with large values of transverse momentum in a high 
energy hadron collision has long been regarded as an incisive probe of short
distance dynamics.  The dominance of the Compton subprocess 
$q + g \rightarrow \gamma + q$ at both leading- and at next-to-leading 
orders\cite{oldboys} in perturbative quantum chromodynamics [QCD] makes 
spin-averaged inclusive prompt photon production the reaction of choice for 
investigations of the magnitude and 
Bjorken $x$ dependence of the gluon density of the incident hadrons, 
$G(x,\mu)$.  The Compton subprocess also dominates the dynamics in 
longitudinally polarized proton-proton 
reactions\cite{bergerqiu,gordonvogelang} as long as the polarized gluon density 
$\Delta G(x,\mu)$ is not too small.  As a result, two-spin measurements of 
inclusive prompt photon production in polarized $pp$ scattering should 
constrain the size, {\it {sign}}, and Bjorken $x$ dependence of 
$\Delta G(x,\mu)$.  

In previous papers we investigated the associated production of a prompt
photon along with a charm quark\cite{bergergordon,bailey}, 
$p +\bar{p}\rightarrow \gamma + c + X$, in spin-averaged reactions at 
high energy.  We showed that 
collider data from this two-particle inclusive reaction at large values of 
transverse momentum may be used to measure the spin-averaged charm quark 
density in the nucleon.  In this paper, we extend our previous work by 
including a full next-to-leading order treatment of the longitudinal 
spin dependence.  We specialize to $p +p \rightarrow \gamma + c + X$ at the 
energies of the Brookhaven Relativistic Heavy Ion Collider (RHIC) facility.  

For values of the transverse momentum $p_T^c$ of the 
charm quark much larger than the mass $m_c$ of the quark, only one 
{\it {direct}} hard scattering subprocess contributes in leading order: 
the quark-gluon Compton subprocess $g c \rightarrow \gamma c$.  The initial 
charm quark and the initial gluon are constituents of the initial hadrons.  In 
addition, there is a leading order {\it {fragmentation}} process in which the
photon is produced from quark or gluon fragmentation, e.g.,
$g g \rightarrow c \bar{c}$ followed by $\bar{c} \rightarrow \gamma X$, or
$q c \rightarrow  q c$ followed by $q \rightarrow \gamma$.
At next-to-leading order in QCD, several subprocesses contribute to the
$\gamma + c +X$ final state: $gc \rightarrow gc\gamma$,
$g g \rightarrow c \bar{c} \gamma$,
$q \bar{q} \rightarrow c \bar{c} \gamma$,
$q c \rightarrow  q c \gamma$,
$\bar{q} c \rightarrow \bar{q} c \gamma$,
$c \bar{c} \rightarrow c \bar{c} \gamma$, and
$c c \rightarrow c c \gamma$.
A full next-to-leading order calculation requires the computation of the 
hard-scattering matrix elements for these two-to-three particle production 
processes as well as the one-loop $O(\alpha_s^2)$ corrections to the lowest 
order subprocess $g c \rightarrow \gamma c$.   

We are interested ultimately in the fully differential two-particle
inclusive cross section,  $E_\gamma E_cd\sigma/d^3p_\gamma d^3p_c$, where
$(E,p)$ represents the four-vector momentum of the $\gamma$ or $c$ quark.  For 
each contributing subprocess, this calculation requires integration over the
momentum of the unobserved final parton in the two-to-three particle 
subprocesses
($g$, $\bar{c},q$, or $\bar{q}$).  Collinear singularities must be handled
analytically by dimensional regularization and absorbed into parton momentum
densities or fragmentation functions.  In the theoretical analysis reported 
here, a combination of analytic and Monte Carlo integration methods is used to 
perform phase-space integrations over unobserved final-state partons and the
momenta of the initial partons.  This approach facilitates imposition of 
photon isolation restrictions and other selections of relevance in experiments. 
We work in the massless approximation, $m_c = 0$.  To warrant use of 
perturbation theory and the massless approximation, we
limit our considerations to values of transverse momenta of the photon
and charm quark $p^{\gamma,c}_{T} > 5$ GeV.  

In the lowest order direct subprocess, $g c \rightarrow \gamma c$, the prompt 
photon emerges in isolation from the only other particle in the hard 
scattering, the charm quark.  Long-distance quark-to-photon
and gluon-to-photon fragmentation processes have been
emphasized theoretically\cite{field} and parametrized phenomenologically in
leading order\cite{frag1}, and evolved in next-to-leading
order\cite{frag2,frag3}. These terms may account for more than half of the
calculated inclusive single photon cross section at modest values of transverse
momentum at collider energies.  Photons originating through
fragmentation are likely to emerge in the neighborhood of associated hadrons.
An experimental
isolation restriction is needed before a clean identification can be made of
the photon and a measurement made of its momentum.  Isolation reduces the size
of the observed fragmentation contribution.  Photon
isolation complicates the theoretical interpretation of results, however, since
it threatens to upset the cancellation of infra-red divergences in perturbation
theory\cite{berqiu}. In this paper, we include the fragmentation contributions, 
and we impose isolation requirements through our Monte Carlo 
method.  

A combination of analytic and Monte Carlo methods similar to that we employ 
in this paper has been used to carry out next-to-leading order QCD 
calculations of other processes including unpolarized and polarized inclusive 
prompt photon\cite{baer1,gordon}, unpolarized and polarized prompt
photon plus jet production \cite{owens,gordon2} and unpolarized and polarized photon pair 
\cite{bailey1,cg} production in hadron 
collisions, single \cite{baer2} and pair production of heavy gauge 
bosons \cite{ohnemus}, and in our earlier work on spin-averaged $\gamma c$ 
production \cite{bailey}.  The combination of analytic and Monte Carlo 
techniques used here to perform the phase space integrals is documented and 
described in detail
elsewhere~\cite{baer1,bailey1,baer2,ohnemus,bailey,cg} with some details
specific to the polarized case discussed in Ref.\cite{cg}.   
We refer readers to those papers for further details.   

After theoretical expressions are derived in perturbative QCD that 
relate the spin-dependent cross section at the hadron level to spin-dependent 
partonic hard-scattering matrix elements and polarized parton densities, we 
must adopt models for spin-dependent parton densities in order to obtain 
illustrative numerical expectations.   For the spin-dependent gluon density 
that we need, we use the three different parametrizations of $\Delta G(x,\mu)$ 
suggested in Ref.~\cite{GS}.  We generate the spin dependent 
polarized charm quark densities $\Delta c(x,\mu)$ perturbatively from these 
polarized gluon densities, beginning with the assumption that 
$\Delta c(x,\mu_o) = 0$ at the starting value for evolution $\mu_o = 1.5$ GeV.  
Our polarized gluon and charm quark densities satisfy expected positivity 
constraints.

We present spin-averaged cross sections differential in the transverse 
momenta and rapidities of the photon and charm quark in $pp$ collisions at the 
center-of-mass energy $\sqrt{S}=200$ GeV typical of the Brookhaven RHIC 
collider.  Our results on the longitudinal 
spin dependence are expressed in terms of the 
two-spin longitudinal asymmetry $A_{LL}$, defined by 
\begin{equation}
A_{LL} = \frac{\sigma^{\gamma,c}(+,+)-\sigma^{\gamma,c}(+,-)}
{\sigma^{\gamma,c}(+,+)+\sigma^{\gamma,c}(+,-)},
\end{equation}
where $+,-$ denote the helicities of the incoming protons. 

In Section II, we describe the next-to-leading order calculation of the 
spin-dependent cross section in perturbative QCD.  Parametrizations of the 
spin-dependent gluon and charm quark densities are discussed in Sec.~III. 
Differential cross sections and other numerical results  
are presented in Section IV.  Summary remarks are collected in
Section V.  An appendix is included in which we derive analytic expressions 
for some of the parton level spin-dependent cross sections.  

\section{Contributions Through Next-to-Leading Order }

The two-particle inclusive hadron reaction  
$h_1 + h_2 \rightarrow \gamma + c + X$ 
proceeds through partonic hard-scattering processes involving
initial-state light quarks $q$ and gluons $g$. In lowest-order QCD, at
${\cal O}(\alpha_s)$, the only {\it {direct}} partonic subprocess is $c + g
\rightarrow \gamma + c$.  In 
addition, there is a leading order {\it {fragmentation}} process in which the
photon is produced from quark or gluon fragmentation, e.g.,
$g g \rightarrow c \bar{c}$ followed by $\bar{c} \rightarrow \gamma X$, or
$q c \rightarrow  q c$ followed by $q \rightarrow \gamma$.  Calculations 
of the cross section at order ${\cal O}(\alpha_s^2)$ involve virtual gluon loop
corrections to the ${\cal O}(\alpha_s)$ direct subprocess 
as well as real gluon radiation contributions from a wide range of 
$2 \rightarrow 3$ parton subprocesses (of which some examples are shown in 
Fig.~1(c).)

The full set of three-body final-state subprocesses is:
\begin{mathletters}\label{eq:1}
\begin{eqnarray}
g &+ c \rightarrow g + c + \gamma\label{eq:11}\\
g &+ g \rightarrow c +\bar{c} + \gamma\label{eq:12}\\
q &+ \bar{q} \rightarrow c +\bar{c} + \gamma\label{eq:13}\\
q &+ c \rightarrow  q + c + \gamma\label{eq:14}\\
\bar{q} &+ c \rightarrow \bar{q} + c + \gamma\label{eq:15}\\
c &+ \bar{c} \rightarrow c + \bar{c} + \gamma\label{eq:16}\\
c &+ c \rightarrow c + c + \gamma\label{eq:17}
\end{eqnarray}
\end{mathletters} 

The physical cross section is obtained through the factorization theorem,
\begin{equation}
 \frac{d^2\sigma_{h_1h_2}^{\gamma,c}}{dp_T^cdp_T^{\gamma}dy_cdy_{\gamma}d\phi} 
 \sim \sum_{ij} \int dx_1 dx_2
 f^i_{h_1}(x_1,\mu_f) f^j_{h_2}(x_2,\mu_f) 
 \frac{sd^2\hat{\sigma}_{ij}^{\gamma,c}}
 {dtdu}(s,p_T,y,\phi;\mu_f).
 \label{eq:gcspin1}
\end{equation}
It depends on the hadronic center-of-mass energy $S$, the transverse momenta 
$p_T^c$ and $p_T^{\gamma}$ of the charm quark and photon, the rapidities 
$y_c$ and $y_{\gamma}$, and the relative azimuthal angle $\phi$; $\mu_f$ 
is the factorization scale of the scattering process.  The
usual Mandelstam invariants in the partonic system are defined by $s =
(p_1+p_2)^2,~t = (p_1-p_{\gamma})^2$, and $u = (p_2-p_{\gamma})^2$,
where $p_1$ and $p_2$ are the momenta of the initial state partons and
$p_{\gamma}$ is the momentum of the final photon.  The indices $ij$ label 
the initial parton channels whose contributions are added incoherently to 
yield the total physical cross section. The spin-averaged parton densities are 
denoted $f_h(x,\mu_f)$.  The partonic hard-scattering cross section  
$\hat\sigma_{ij}^{\gamma,c}(s,p_T,y,\phi;\mu_f)$ is obtained commonly
from fixed-order QCD calculations through
\begin{equation}
 \frac{d^2\hat{\sigma}_{ij}^{\gamma,c}}{dtdu} =
   \alpha_s  (\mu^2) \frac{d^2\hat{\sigma}_{ij}^{\gamma,c,(a)}}{dtdu}
 + \alpha_s^2(\mu^2) \frac{d^2\hat{\sigma}_{ij}^{\gamma,c,(b)}}{dtdu}
 + \alpha_s^2(\mu^2) \frac{d^2\hat{\sigma}_{ij}^{\gamma,c,(c)}}{dtdu}
 + {\cal O} (\alpha_s^3).
 \label{eq:gcspin2}
\end{equation}
The tree, virtual loop, and real emission contributions are labeled
(a), (b), and (c).  The parameter $\mu$ is the renormalization scale. 

The longitudinal spin-dependent cross section for 
$h_1 + h_2 \rightarrow \gamma + c+ X$ is very similar to 
Eq.~(\ref{eq:gcspin1}):  
\begin{equation}
 \frac{d^2\Delta 
\sigma_{h_1h_2}^{\gamma,c}}{dp_T^cdp_T^{\gamma}dy_cdy_{\gamma}d\phi}
\sim \sum_{ij} \int dx_1 dx_2
 \Delta f^i_{h_1}(x_1,\mu_f) \Delta f^j_{h_2}(x_2,\mu_f) 
 \frac{sd^2 \Delta \hat{\sigma}_{ij}^{\gamma,c}}
 {dtdu}(s,p_T,y,\phi;\mu_f).
 \label{eq:gcspin3}
\end{equation}

The polarized parton densities are defined by
\begin{equation}
\Delta f^i_h(x,\mu_f)=f^i_{h,+}(x,\mu_f)-f^i_{h,-}(x,\mu_f), 
\end{equation}
where $f^i_{h\pm}(x,\mu_f)$ is the distribution of parton of type $i$
with positive $(+)$ or negative $(-)$ helicity in hadron $h$. 
Likewise, the polarized partonic cross section  
$\Delta \hat {\sigma}^{\gamma,c}$ is 
defined by 
\begin{equation}
\Delta \hat{\sigma}^{\gamma,c}=\hat {\sigma}^{\gamma,c}(+,+)-
\hat {\sigma}^{\gamma,c}(+,-), 
\end{equation}
with $+,-$ denoting the helicities of the incoming partons. 

\subsection{Leading order contributions}

In leading order in perturbative QCD, only one {\it{direct}} subprocess
contributes to the hard-scattering cross section, the QCD Compton process
$c g\rightarrow \gamma c$,
unlike the case for single inclusive prompt photon production, where the
annihilation process $q\bar{q}\rightarrow \gamma g$ also contributes.
Since the leading order direct partonic subprocess has a two-body final state,
the photon and $c$ quark are produced with balancing transverse
momenta.  In addition, there are effectively leading-order contributions in 
which the photon is produced by fragmentation from a final-state parton. These 
are
\begin{eqnarray}
c+g&\rightarrow& g+c \nonumber \\
g+g&\rightarrow& c+\bar{c} \nonumber \\
c+q&\rightarrow& c+q \nonumber \\
c+\bar{q}&\rightarrow&c+\bar{q} \nonumber \\
c+c&\rightarrow& c+c \nonumber \\
c+\bar{c}&\rightarrow & c+\bar{c} \nonumber \\
q+\bar{q}&\rightarrow & c+\bar{c}.
\label{eq:fragproc}
\end{eqnarray}
If the photon is to be isolated from the observed charm quark, it arises from 
fragmentation of the gluon $g$ and the non-charm quark $q$, respectively,
in the cases of the first, third and fourth processes.  In the
other cases it is produced by fragmentation of one of the (anti)charm quarks.

In a fully consistent next-to-leading order calculation, one should calculate
the subprocesses in Eq.~(\ref{eq:fragproc}) to $O(\alpha_s^3)$, since the photon
fragmentation functions that are convoluted with the hard subprocess
cross sections are of $O(\alpha_{em}/\alpha_s)$.  For simplicity, we
include them in $O(\alpha_s^2)$ only.  In fact, next-to-leading order
fragmentation contributions to single prompt photon production have been
included only once before\cite{gorvogel2,gordon}. We expect the next-to-leading order
corrections to the fragmentation contributions to be insignificant numerically
especially after isolation cuts are imposed. This expectation is
confirmed by a study performed in Ref.\cite{gordonx} on inclusive prompt
photon production in the polarized case. It was also shown in
Ref.\cite{gordonx} that predictions for the asymmetries are hardly
affected by inclusion of the fragmentation contributions either before
or after isolation of the photon.   

\subsection{Next-to-leading order contributions}

There are two classes of contributions in next-to-leading order.  There
are the virtual gluon exchange corrections to the lowest order process, 
$c g\rightarrow \gamma c$. Examples are
shown in Fig.1(b). These amplitudes interfere with the Born amplitudes and
contribute at $O(\alpha_{em}\alpha_s^2)$. They were calculated
twice before in the spin-averaged case\cite{aurenche,gorvogel}.  
At next-to-leading order there are also
three-body final-state contributions, listed in Eq.~(\ref{eq:1}). Both the
spin dependent and spin-averaged virtual loop and three-body 
matrix elements are taken from Ref.~\cite{gorvogel}, where 
they are calculated for single inclusive prompt photon production.    
 
In Ref.\cite{gorvogel} the original 'tHooft-Veltman-Breitenlohner-Maison
(TVBM) \cite{thooft,maison} scheme was used to treat the 4-dimensional 
$\gamma_5$ matrix and antisymmetric $\epsilon_{\mu\nu\rho\sigma}$
tensor in $n$ dimensions. These objects arise when the traces are taken
to calculate the helicity dependent matrix elements. To project onto
definite helicity states, $h$, for (anti-) quarks and $\lambda$ for
gluons, the relations
\begin{equation}
u(p,h)\bar{u}(p,h)=\frac{1}{2}\gamma^{\mu}(p)(1-h \gamma_5)
\end{equation}
and  
\begin{equation}
\epsilon_{\mu}(p,\lambda)\epsilon_{\nu}^*(p,\lambda)=
\frac{1}{2} \left[-g_{\mu\nu}+i\lambda\epsilon_{\mu\nu\rho\sigma}
\frac{p^{\rho}p'^{\sigma}}{p.p'}\right]
\end{equation}
are used. 
In the TVBM scheme, $n$-dimensional Minkowski space is divided into
a $4$-dimensional and a $(n-4)$-dimensional part.  Any vector $k$
has a $4$-dimensional, $\hat{\hat{k}}$, and a ($n-4$)-dimensional part,
$\hat{k}$. This means that the $n$ dimensional matrix elements
calculated in Ref.\cite{gorvogel} contain products of these
($n-4$)-dimensional `hat' momenta. In Ref.\cite{gorvogel} it was shown
that these contributions are non-zero in the collinear limit only, and thus
in our case they are included in the hard collinear integrals given in
the appendix. A detailed account of all aspects of the phase space
integration will be given elsewhere \cite{gordon3}. It should be
mentioned here that the matrix elements were also calculated in
Ref.~\cite{conto} in a scheme different from the TVBM scheme outlined here.
Their results seem similar to those of Ref.\cite{gorvogel}, but a
detailed comparison of the matrix elements has not yet been made.  

The main task of our calculation is to integrate the three-body matrix elements
over the phase space of
the unobserved particle in the final state. The situation here is
different from the standard case of single inclusive particle
production because we wish to
retain as much control as possible over the kinematic variables of a
second particle in the final state, while at the same time integrating
over enough of the phase space to ensure cancellation of all infrared
and collinear divergences, inherent when massless
particles are assumed.  All the processes of Eq.~(\ref{eq:1}), except the 
first, involve collinear singularities but no soft singularities. The soft 
and collinear singularities are first 
exposed by integrating the three-body phase space in $n$-dimensions over
the soft and
collinear regions of phase space. These soft and collinear regions are 
separated from the rest of the
three body phase space by introducing soft and collinear cut-off
parameters $\delta_s$ and $\delta_c$.   
The collinear and soft singularities are cancelled and factored in the
$\overline{MS}$ scheme as explained in Ref.~\cite{bailey}.

At $O(\alpha ^2_s)$ there are, in addition, fragmentation processes
in which the hard-scattering two-particle final-state subprocesses
\begin{eqnarray}
c+g &\rightarrow& \gamma+ c \nonumber \\
c+\bar{c}&\rightarrow& \gamma+g \nonumber \\
q+\bar{q}&\rightarrow& \gamma +g
\end{eqnarray}
are followed by fragmentation processes $c\rightarrow c X$, in the case of
the first subprocess, and
$g\rightarrow c X$ in the cases of the last two. These should be included
because we have factored the collinear singularities in the
corresponding three-body final-state processes into non-perturbative 
fragmentation functions for production of a charm quark from a particular 
parton. As a first approximation, we estimate these fragmentation functions by
\begin{eqnarray}
D_{c/c}(z,\mu^2)&=&\frac{\alpha_s(\mu^2)}{2\pi}P_{qq}(z),
\label{eq:cfrag}
\end{eqnarray}
and
\begin{eqnarray}
D_{c/g}(z,\mu^2)&=&\frac{\alpha_s(\mu^2)}{2\pi}P_{qg}(z),
\label{eq:gfrag}
\end{eqnarray}
where $P_{ij}(z)$ are the lowest order splitting functions for parton
$j$ into parton $i$ \cite{altar}; and $\alpha_s(\mu^2)$ is the strong
coupling strength. 

For completeness we have gathered some of the main ingredients used in
calculating the polarized cross section in the appendix. Similar results
for the unpolarized case can be found in Ref.~\cite{bailey}. Further details 
and a complete list of the $4$-dimensional three-body final-state matrix 
elements will be given in Ref.\cite{gordon3}.  

\section{Polarized Parton Densities }

Our spin-dependent gluon density and our spin-dependent up, down, and 
strange quark densities (and their antiquark components) are adapted from the 
set of polarized parton densities published by Gehrmann and Stirling 
(GS)~\cite{GS}. 
However, this set of spin-dependent densities, as well as other 
recent parametrizations available for general use, does not include a charm 
contribution \cite{ggr}.  At the values of squared four-momentum transfer, 
$Q^2$, at which spin-dependent deep-inelastic scattering data have been 
obtained, the charm contribution is expected to be very small.  Fits to the 
data are thus insensitive to the charm contribution. The situation is somewhat 
different in the spin-averaged case, and mechanisms for inclusion of charm 
are in general use.

In particular, a charm component is included in the CTEQ4M spin-averaged 
parametrization~\cite{cteq} at a scale $\mu_o=m_c$, where $m_c\sim 1.5$ GeV is 
the charm quark mass.  At this threshold scale the charm quark density is 
assumed to have zero value, $c(x,\mu_o)=0$.  Non-zero charm quark and 
antiquark densities are generated at larger values of $\mu$ via the DGLAP 
perturbative QCD
evolution equations through gluon splitting into $c\bar{c}$ pairs.  Since we 
use the CTEQ4M distributions~\cite{cteq} as our spin-averaged parton 
densities, it is reasonable to use a similar mechanism to generate the 
spin-dependent charm densities, with spin-dependent Altarelli-Parisi splitting 
functions used in the evolution.  

The evolution mechanism outlined above implies that the spin-dependent charm 
quark density will depend strongly on the assumed value of the spin-dependent 
gluon density.  Uncertainty in knowledge of the gluon density will 
propagate to the charm density.  The current deep 
inelastic scattering data do not constrain the polarized gluon density
very tightly, and most groups present more than one plausible parametrization.  
Gehrmann and Stirling~\cite{GS} present three such parametrizations, 
labelled GSA, GSB, and GSC.  In our parametrizations of the spin-dependent 
densities we begin with the GS parametrizations as our initial set at a 
scale $\mu_o=m_c$, and we evolve them to obtain spin-dependent gluon and quark 
densities at greater values of $\mu$, including a spin-dependent charm quark 
density.  We use next-to-leading order DGLAP 
evolution equations with the number of flavors, $n_f=4$.  Details of the 
$x$-space evolution will be reported elsewhere~Ref.~\cite{gr}.  Our approach is 
to adopt the GS polarized parton densities as our starting values and to extend
them to include charm.

In Fig.~2(a), we display the ratio of our spin-dependent and spin-independent 
gluon densities, $\Delta G(x,\mu)/G(x,\mu)$, at the hard scale $\mu^2 = 100$ 
GeV$^2$ for the GSA, GSB, and GSC choices.  In the GSA and GSB sets, $\Delta 
G(x,\mu_o)$ is positive for all $x$, whereas in the GSC set 
$\Delta G(x,\mu_o)$ changes sign.  After evolution to $\mu^2 = 100$ GeV$^2$, 
$\Delta G(x,\mu)$ remains positive for essentially all $x$ in all three sets, 
but its magnitude is small in the GSB and GSC sets.  In Fig.~2(b), we show 
the spin-dependent charm quark distributions.  Positivity is 
well satisfied, as it is for $\Delta G(x,\mu)/G(x,\mu)$.  The value of 
$\Delta c(x,\mu)/c(x,\mu)$ is reasonably substantial in the GSA case, 
reflecting the large size of $\Delta G(x,\mu)$ from which $\Delta c(x,\mu)$ 
is generated.  The size of $\Delta c(x,\mu)/c(x,\mu)$ is correspondingly smaller 
in the GSB and GSC cases, hovering near zero in the GSC case.  

As mentioned, our charm quark densities depend on the choice made for the 
gluon densities.  In addition, we assume that there is no intrinsic charm 
quark density.  If an intrinsic density were adopted, one might begin with 
non-zero charm and anti-charm densities at $\mu_o$ whose $x$-dependences need 
not be assumed identical, in both the spin-averaged and spin-dependent cases.  
In a sense, our assumed forms for the spin-dependent charm quark densities 
represent the most conservative possibilities, and one should be alert 
experimentally to more interesting outcomes.  

\section{Numerical Results}

In this section we present and discuss spin-averaged differential
cross sections and two-spin longitudinal asymmetries for the joint production 
of a charm quark and a 
photon at large values of transverse momentum. All results are displayed
for $pp$ collisions at the center-of-mass
energy $\sqrt{S}=200$ GeV typical of the Brookhaven RHIC collider.
To obtain the spin-averaged differential cross sections presented in this 
paper, we convolute our hard-scattering matrix elements with the CTEQ4M parton 
densities\cite{cteq}.  We use the standard two-loop formula for the strong 
coupling strength with four massless flavors of quarks.  We set 
$\Lambda^{(5)}_{QCD}=0.202$ GeV (the CTEQ4M value).  The spin-dependent parton 
densities are described in Sec.~III.  

Very similar differential distributions may be obtained if other parton sets 
are used instead, with quantitative differences reflecting differences among 
gluon and charm quark densities in the different 
sets\cite{bergergordon}.  We set the renormalization, factorization, and 
fragmentation scales to a common value $\mu = p_T^{\gamma}$ in most
of our calculations.  Since there are two observed particles in the final state,
the charm quark and the photon, both of whose transverse momenta are
large, an alternative choice might be $\mu = p_T^c$ or some function
of $p_T^{\gamma}$ and $p_T^c$.  The results of our calculations
show that the magnitudes of $p_T^{\gamma}$ and $p_T^c$ tend to be 
comparable and that dependence of the cross sections on $\mu$ is slight.  
Therefore, choices of $\mu$ different from $\mu = p_T^{\gamma}$ should not 
produce significantly different answers, and we have verified this 
supposition in representative cases.

In collider experiments a photon is observed and its momentum is well
measured only when the photon is isolated from neighboring hadrons.  In
our calculation, we impose isolation in terms of the cone variable $R$:  
\begin{equation}
\sqrt{(\Delta y)^2 + (\Delta \phi)^2} \leq R.     \label{eq:Rdef}
\end{equation}
In Eq.~(\ref{eq:Rdef}), $\Delta y$ ($\Delta \phi$) is the difference between 
the rapidity (azimuthal angle in the transverse plane) 
of the photon and that of any parton in the final state.  The photon is said
to be isolated in a cone of size $R$ if the ratio of the hadronic energy in the 
cone and the transverse momentum of the photon does not exceed 
$\epsilon = {\rm 2 GeV}/p_T^{\gamma}$. We show distributions
for the choice $R = 0.7$.  

In Fig.~3(a) we show the spin-averaged differential cross section as a 
function of the
transverse momentum of the photon $p_T^{\gamma}$, having restricted the
transverse momentum of the charm quark to the range $p^c_T\geq 5$ GeV.  The 
rapidities of the charm quark and photon are restricted to the central
region, $-1\leq y^{\gamma,c}\leq 1$ in order to mimic the central region 
coverage of the major detectors at RHIC.  The solid curve shows our prediction 
when contributions are included from all subprocesses through $O(\alpha ^2_s)$.
The dominance of the $cg$ subprocess is illustrated by our dashed curve.  This 
dominance is the basis for the statement that the spin-averaged cross 
section at collider energies can be used to determine the magnitude and 
Bjorken $x$ dependence of the spin-averaged charm quark 
density\cite{bergergordon,bailey}.  

In Figs.~3(b) and (c), we present the two-spin longitudinal asymmetries, 
$A_{LL}$, as a function of  $p_T^{\gamma}$, for the same kinematic selections 
as made for Fig.~3(a).  Results are shown for three choices of the 
polarized gluon density (and, correspondingly, for the polarized charm 
quark density).  The asymmetry becomes sizeable for large enough 
$p_T^{\gamma}$ only in the case of the GSA parton set.  

As noted above the $cg$ subprocess dominates the {\it{spin-averged}} cross 
section. It is interesting and important to inquire whether this dominance 
persists in the spin-dependent situation.  The dot-dashed curve in Fig.~3(b) 
shows the contribution to the asymmetry from the polarized $cg$ subprocesses 
in the case of the GSA set.  It is positive for all $p_T^{\gamma}$.  At small 
$p_T^{\gamma}$, the net asymmetry is driven negative by a large contribution 
from the $gg$ subprocess.  For this GSA set, we see that once it becomes 
sizable (e.g., 5\% or more), the total asymmetry from all subprocesses is 
dominated by the large contribution from the $cg$ subprocess.  In Fig.~3(c), 
using a more expanded scale, we replot the overall asymmetry 
for the GSC choice of the polarized gluon density, and we also show the 
contribution from the $cg$ subprocess alone.  In this case, the overall 
asymmetry $A_{LL}$ itself is small, and the contribution from the 
$cg$ subprocess cannot be said to dominate the answer.    

In interpreting the results presented in Figs.~3(b) and (c), we begin with the 
supposition that the polarized gluon density will have been determined from 
data on inclusive prompt photon production.  The question to pose is 
whether asymmetries of the type shown in Figs.~3(b) and (c) could shed 
light on the polarization of the charm quark density.  If a large asymmetry 
is measured, similar to that expected in the GSA case at the larger values 
of $p_T^{\gamma}$, Fig.~3(b) shows that 
the answer is dominated by the $cg$ contribution, and the data will serve to 
constrain $\Delta c(x,\mu)$.  On the other hand, if $\Delta G(x,\mu)$ is small 
and a small asymmetry is measured, such as shown in Fig.~3(c), one will not be 
able to conclude which of the subprocesses is principally responsible, and no 
information could be adduced about $\Delta c(x,\mu)$.  

In Fig. 4 we show distributions in the rapidity of the charm quark, $y^c$,
for $-1\leq y^\gamma\leq 1$, $p^c_T\geq 5$ GeV, and 
4 GeV $\leq p_T^\gamma\leq$ 50 GeV.  The spin-averaged distribution in 
$y^c$, shown in Fig.~4(a) is fairly broad, with 
full-width at half-maximum of nearly 3 units in rapidity.  The dashed 
curve in Fig.~4(a) shows that the contribution from the $cg$ subprocess is 
dominant.  The asymmetries are shown in Fig.~4(b) as functions of $y^c$.  
They are small for all three choices of the the polarized gluon density.
These results are consistent with those shown in Fig.~3(b), 
although they might not appear to be at first glance.  One must bear in mind 
that it is the region of relatively small $p_T^\gamma$ that dominates the 
integral over $p_T^\gamma$.  The asymmetries shown in Fig.~3(b) are small 
at modest values of $p_T^\gamma$.  In the GSA case, the total asymmetry at 
relatively small values of $p_T^\gamma$ is opposite in sign to the positive 
contribution from the $cg$ subprocess.   

The structure of the QCD hard-scattering matrix element produces 
{\it {positive}} correlations in rapidity\cite{elbcor} at collider energies. 
To examine correlations more precisely, we study the spin-averaged cross 
section and the asymmetry as functions of the difference of the rapidities of 
the photon and charm quark.  Results are shown in Fig.~5.

\section{Summary and Discussion}

In this paper we present the results of a calculation of the 
longitudinal spin-dependence of 
the inclusive production of a prompt photon in association with a charm quark at
large values of transverse momentum.  This analysis is done at
next-to-leading order in perturbative QCD.  We employ a combination of 
analytic and Monte Carlo integration methods in which infrared and 
collinear singularities of the next-to-leading order matrix elements are
handled properly.  We provide differential cross sections and polarization 
asymmetries as functions of transverse momenta and rapidity, including
photon isolation restrictions, that may be useful for estimating the 
feasibility of measurements of the spin-averaged and spin-dependent cross 
sections in future experiments at RHIC collider energies.  We show 
that the study of two-particle inclusive spin-averaged distributions, with 
specification
of the momentum variables of both the final prompt photon and the final heavy
quark, tests correlations inherent in the QCD matrix elements\cite{elbcor} 
and should provide a means for measuring the charm quark density in the 
nucleon\cite{bergergordon}.  In the spin-dependent case, significant values 
of $A_{LL}$ (i.e., greater than 5 \%) may be expected for $p_T^{\gamma} > 
15$ GeV if the polarized gluon density $\Delta G(x,\mu)$ is as large as that 
in the GSA set of polarized parton densities.  If so, the data on associated 
production could be used to determine the polarization of the charm quark 
density in the nucleon.  On the other hand, for small $\Delta G(x,\mu)$, 
dominance of the $cg$ subprocess is lost, and $\Delta c(x,\mu)$ is 
inaccessible.  

Our spin-averaged and polarized charm quark densities are generated 
perturbatively from gluon splitting into charm quark and charm antiquark 
pairs.  They depend entirely on the choice made for the gluon densities: a 
small polarized gluon density leads to a small polarized charm density. 
Reality may well be different. For example, there could be a non-perturbative 
intrinsic charm quark component.  In a sense, our assumed forms for the 
spin-averaged and spin-dependent charm quark densities represent the most 
conservative possibilities.  

In a typical experiment, the momentum
of the quark may be inferred from the momentum of prompt lepton decay products
or the momentum of charm mesons, such as $D^*$'s.  Alternatively, our
distributions in $p^c_{T}$ may be convoluted with charm quark
fragmentation functions, deduced from, e.g., $e^+e^-$ annihilation
data, to provide distributions for the prompt leptons or $D^*$'s.

\section{Acknowledgments}

The work at Argonne National Laboratory was supported by the US Department of
Energy, Division of High Energy Physics, Contract number W-31-109-ENG-38.
\pagebreak

\appendix

\section{Analytic Expressions}

In order to make this paper reasonably self-contained, we collect in
this appendix all the formulae we use in the calculation of the
polarized cross section.  We label
the momenta for the generic three-body final-state process by
\begin{equation}
p_1+p_2\rightarrow p_3+p_4+p_5 ,
\label{eq:aone}
\end{equation}
where $p_3$ denotes the photon and $p_4$ denotes the observed charm quark, 
and we define the usual Mandelstam invariants
\begin{eqnarray}
\hat{t}&=&(p_1-p_3)^2 \nonumber \\
\hat{u}&=&(p_2-p_3)^2 \nonumber \\
\hat{s}&=&(p_1+p_2)^2.
\label{eq:atwo}
\end{eqnarray}
We express the two-body final-state cross sections in terms of the scaled
variable $v$, where
\begin{equation}
v=1+\frac{\hat{t}}{\hat{s}}.
\label{eq:athree}
\end{equation}

\subsection{Two-body contributions}

The effective two-body contribution includes the $O(\alpha ^2_s)$ virtual
gluon-exchange loop
contributions and the soft and/or collinear parts of the three-body
contributions (this remark applies to initial-state collinear contributions
only, as explained later).  After all soft pole singularities are 
cancelled and all collinear pole singularities are factored,
the effective two-body contribution is expressed as
\begin{eqnarray}
\Delta\sigma_{2body}(A+B\rightarrow\gamma+c+X)&=& \int dv dx_1 dx_2\left[
\frac{d\Delta \sigma_{coll}^{cg\rightarrow \gamma c}}{dv} \right. \nonumber \\
&+&\left. \Delta f^A_g(x_1,M^2)\Delta f^B_c(x_2,M^2)\frac{d\Delta\sigma^{HO}}
{dv}(cg\rightarrow
\gamma c)\right],
\label{eq:afour}
\end{eqnarray}
plus terms in which the beam and target are interchanged. We use subscript 
$c$ to label the charm (or anti-charm) quark. 

We define 
\begin{equation}
\Delta T_{cg}=\frac{v(2-v)}{1-v} ,
\end{equation}
\label{eq:afive}
and $v_1=1-v$.  In Eq.~(\ref{eq:afour}),
\begin{eqnarray}
\frac{d\Delta\sigma^{HO}}{dv}(cg\rightarrow \gamma c)&=&\frac{\pi\alpha_{em}
\alpha_s e_c^2}{\hat{s}N_C}
\left(\Delta T_{cg}+\frac{\alpha_s}{2\pi}
\left[\frac{1}{2}\left(-14 C_F \Delta T_{cg} + 2 (2 C_F + N_c) 
\ln^2\delta_s \Delta
T_{cg} \right. \right. \right. \nonumber \\
&-& \left. \left. \left.  \frac{2}{3} N_F \ln\frac{\hat{s}}{M^2} \Delta T_{cg}
 + C_F (3 + 4 \ln\delta_s)\left(
\ln\frac{\hat{s}}{M^2}+\ln\frac{\hat{s}}{M''^2}
\right) \Delta T_{cg}
\right. \right. \right. \nonumber \\
&+& \left. \left. \left. \frac{N_c}{3}(11 + 12 \ln \delta_s)
\ln \frac{\hat{s}}{M^2} \Delta
T_{cg} + \frac{1}{3} (11 N_c - 2 N_F) \ln\frac{\hat{s}}{\mu^2} \Delta
T_{cg} \right. \right. \right. \nonumber \\
&+& \left. \left. \left.  (2 C_F - N_c)\ln^2 v\Delta T_{cg} + 
4\ln\delta_s (N_c \ln v_1 + 
2 C_F \ln v -  N_c \ln v) \Delta T_{cg}  \right. \right. \right.
\nonumber \\
&+& \left. \left. \left.  2 (2 C_F - N_c) \ln v_1 \ln v
\Delta T_{cg} - 2 C_F \pi^2 \frac{(3 - 4 v - v^2)}{3 v_1} \right.
\right. \right. \nonumber \\
&+& \left. \left. \left. N_c \pi^2 \frac{(3 - 2 v - 2 v^2)}{3 v_1} + 
      \ln^2 v_1 \left( N_c (1 + v) - 2 C_F \frac{(1 - 2 v)}{v_1} \right)
\right. \right. \right. \nonumber \\ 
&+& \left. \left. \left. \ln v_1 \left( 2 N_c + 
2 C_F \frac{(1 + 2 v)}{v_1} \right) + 
2 (2 C_F - N_c) \Delta T_{cg} {\rm Li}_2(v_1) \right. \right. \right. \nonumber
\\
&+& \left. \left. \left. 2 N_c \Delta T_{cg} {\rm Li}_2(v)
\frac{}{}\right) \right] \right).
\label{eq:asix}
\end{eqnarray}

The scales $M$ and $M''$ are the factorization and fragmentation scales,
respectively, on the initial
parton and final-state charm quark legs, and $\mu$ is the 
renormalization scale. $C_F=4/3$ is the
quark-gluon vertex color factor, $N_C=3$ is the number of colors, $e_c$
is the fractional charge of the charm quark, $\delta_s$ and $\delta_c$
are the soft and collinear cutoff parameters introduced in Sec.~II,
${\rm Li}_2(x)$ is the dilogarithm function, and $\alpha_{em}$ is the
electromagnetic coupling constant. 

The remnants of the factorization of the hard collinear singularities
are
\begin{eqnarray}
\frac{d\Delta\sigma^{cg\rightarrow\gamma c}_{coll}}{dv}&=&\frac{\alpha_{em}
\alpha_s^2e_c^2}{2\hat{s}}\frac{1}{N_C}\Delta T_{cg}\left[\Delta f^A_g(x_1,M^2)\left(
\int^{1-\delta_s}_{x_2}\frac{dz}{z}\Delta f^B_c\left(\frac{x_2}{z},M^2\right)
\Delta\tilde{P}_{qq}(z)\right. \right. \nonumber \\
&+&\left. \left. \int^1_{x_2}\frac{dz}{z}\Delta f^B_g\left( \frac{x_2}{z},M^2 \right)
\Delta\tilde{P}_{qg}(z)\right)\right.  \nonumber \\
&+&\left. 
\Delta f^B_c(x_2,M^2)\left(
\int^{1-\delta_s}_{x_1}\frac{dz}{z}\Delta f^A_g\left(\frac{x_1}{z},M^2\right)
\Delta\tilde{P}_{gg}(z)\right. \right. \nonumber \\
&+& \left. \left. \int^1_{x_1}\frac{dz}{z}\Delta f^A_q\left( \frac{x_1}{z},M^2 \right)
\Delta\tilde{P}_{gq}(z)\right) \right].
\label{eq:aseven}
\end{eqnarray}
The last distribution,$\Delta f^A_q(x_1,M^2)$, in Eq.~(\ref{eq:aseven}) implies a sum
over the flavors of quarks from the $c c$, $cq$, and
$c\bar{q}$ initial states. The remaining two processes, $c\bar{c}$ and
$q\bar{q}$, do not have initial state collinear singularities and
thus do not contribute to this part of the cross section. 

The polarized splitting functions $\Delta\tilde{P}_{ij}$, are 
\begin{equation}
\Delta\tilde{P}_{ij}(z)=\Delta P_{ij}(z)\ln\left(\frac{1-z}{z}\delta_c
\frac{\hat{s}}{M^2}\right)-\Delta P'_{ij}(z).
\label{eq:aeight}
\end{equation}
The functions $\Delta P_{ij}(z,\epsilon)$, the spin dependent 
Altarelli-Parisi splitting functions in $4-2\epsilon$ dimensions, are 
\begin{eqnarray}
\Delta P_{qq}(z,\epsilon)&=&C_F\left[\frac{1+z^2}{1-z}+3\epsilon(1-z)\right]\nonumber \\
\Delta P_{qg}(z,\epsilon)&=&\frac{1}{2}\left[(2z-1)-2\epsilon(1-z)\right]\nonumber \\
\Delta P_{gg}(z,\epsilon)&=&2N_C\left[\frac{1}{1-z}-2z+1+2\epsilon(1-z) \right]\nonumber \\
\Delta P_{gq}(z,\epsilon)&=&C_F\left[(2-z)+2\epsilon(1-z)\right].
\label{eq:anine}
\end{eqnarray}
The functions $\Delta P'_{ij}(z)$ are defined by the relation
\begin{equation}
\Delta{P}_{ij}(z,\epsilon)=\Delta P_{ij}(z)+\epsilon\Delta P'_{ij}(z).
\label{eq:aten}
\end{equation}

\subsection{Pseudo-Two-Body Contributions}

Since we are interested in distributions in the kinematic variables of
two final-state partons, the photon and the $c$ quark, we can define
variables that depend on the momenta of both.  An example is
the variable $z$, defined by
\begin{equation}
z=-\frac{p_T^{\gamma}.p_T^c}{|p_T^{\gamma}|^2}.
\end{equation}
Whenever there is a third 
parton in the final state, the distribution in $z$ (or in other analogous 
variables) will differ from a $\delta$~function 
when the third parton carries a finite momentum,
even if it is collinear to one of the other final partons. For this reason we
designate as ``pseudo-two-body contributions" those for which the third parton
is collinear to either the final photon or the charm quark. These
contributions are expressed, respectively, by the equations 
\begin{eqnarray}
\Delta \sigma_{\gamma/coll}&=&\sum_{abcq}\int
\Delta f^A_a(x_1,M^2)\Delta f^B_b(x_2,M^2)\left(\frac{\alpha_{em}}{2\pi}\right) \left[
P_{\gamma q}(z)\ln\left[z(1-z)\delta_c\frac{\hat{s}}{M'^2}\right]-
P'_{\gamma/q}(z)
\right]  \nonumber \\
&\times &\frac{d\hat{\Delta \sigma}}{dv}
(ab\rightarrow cq)dx_1 dx_2 dz dv,
\label{eq:aeleven}
\end{eqnarray}
and
\begin{equation}
\Delta\sigma_{c/coll}=\sum_{abd}\int
\Delta f^A_a(x_1,M^2)\Delta
f^B_b(x_2,M^2)\tilde{P}_{cd}(z,M''^2)\frac{d\Delta \hat{\sigma}}{dv}
(ab\rightarrow \gamma d)dx_1 dx_2 dz dv.
\label{eq:atwelve}
\end{equation}
The functions $P_{\gamma/q}(z)$ and $P'_{\gamma/q}(z)$ are the quark-to-photon
splitting function and $O(\epsilon)$ piece, respectively.  They have the same
form as $P_{gq}$, with the color factor replaced by the square of the quark 
charge.  The scale $M'$ is the fragmentation scale for quark fragmentation 
into a photon.  In Eq.~(\ref{eq:atwelve}),
\begin{equation}
\tilde{P}_{cd}(z,M''^2)=P_{cd}(z)\ln\left[z(1-z)\delta_c\frac{s}{M''^2}\right]
-P'_{cd}(z),
\label{eq:athirteen}
\end{equation}
where $P_{cd}(z)$ represents the splitting functions $P_{qg}(z)$ and
$P_{qq}(z)$.  The unpolarized splitting functions are
\begin{eqnarray}
P_{qq}(z,\epsilon)&=&C_F\left[\frac{1+z^2}{1-z}-\epsilon(1-z)\right]\nonumber \\
P_{qg}(z,\epsilon)&=&\frac{1}{2(1-\epsilon)}\left[z^2+(1-z)^2-\epsilon \right]\nonumber \\
P_{gg}(z,\epsilon)&=&2N_C\left[\frac{z}{1-z}+\frac{1-z}{z}+z(1-z) \right]\nonumber \\
P_{gq}(z,\epsilon)&=&C_F\left[\frac{1+(1-z)^2}{z}-\epsilon z\right].
\label{eq:aninex}
\end{eqnarray}
The functions $P'_{ij}(z)$ are defined by the relation
\begin{equation}
P_{ij}(z,\epsilon)=P_{ij}(z)+\epsilon P'_{ij}(z).
\label{eq:atenx}
\end{equation}
In Eq.~(\ref{eq:aeleven}), $q$ can be a charm quark or a charm anti-quark, or 
a(an) (anti)quark of any flavor in the case of $c q\rightarrow c q$. 
In Eq.~(\ref{eq:atwelve}), $d$ is either a gluon, a charm quark, or a charm
antiquark. Note that the splitting functions $P_{\gamma/q}$ and $P_{cd}$ are
the usual spin averaged and not the spin dependent ones, since the final
state is not polarized. 

These contributions are referred to usually as the remnants of the
factorization of the hard collinear singularities and are regarded as two-body
processes, or as parts of the fragmentation contributions because of their
dependence on the factorization scales.  We prefer to regard
them as pseudo-two-body contributions.  When we examine either the
charm quark or photon momentum distributions, these contributions populate
the same regions of phase space as the other three-body contributions in 
Eq.~(\ref{eq:aseventeen}), unlike the effective two-body contributions.
The pseudo-two-body contributions are usually negative in overall sign due to
the large logarithms of the cut-off parameters, as are the two-body
contributions discussed above. 

\subsection{Photon Fragmentation Contributions}

As mentioned in Sec.~II, we include the quark-to-photon and gluon-to-photon
fragmentation contributions at leading order only.  We convolute the 
$2\rightarrow 2$ hard scattering subprocess cross sections for the processes
listed in Eq.~(\ref{eq:fragproc}) with photon fragmentation functions
$D_{\gamma/i}(z,M'^2)$; $M'$ is the fragmentation scale, the same
scale at which we subtract the collinear singularities on the photon
leg of the three-body processes. The expression for the cross section is
\begin{equation}
\Delta\sigma_{frag/\gamma}=\sum_{abj}\int
\Delta f^A_a(x_1,M^2)\Delta f^B_b(x_2,M^2)D_{\gamma/j}(z,M'^2)
\frac{d\Delta\hat{\sigma}}{dv}
(ab\rightarrow jc)dx_1 dx_2 dz dv.
\label{eq:afourteen}
\end{equation}
The matrix elements for the hard subprocess cross section 
can be found, for example, in Ref.\cite{gorvogel2}.

\subsection{Charm Fragmentation Contributions}

In integrating some of the three-body matrix elements over phase space we
encounter configurations in which the produced charm quark is collinear 
with an anti-charm quark or a gluon in the final state.  These situations lead
to a collinear singularity in the massless approximation.  They occur for 
the processes of Eq.~(\ref{eq:11}), (\ref{eq:13}), and (\ref{eq:16}). We
factor these singularities into a fragmentation function $D_{c/i}(z,M''^2)$ for
parton $i$ to produce a charm quark with momentum fraction $z$. The
contributing subprocess cross sections are 
\begin{eqnarray}
\frac{d\Delta\hat{\sigma}}{dv}(q\bar{q}\rightarrow \gamma
g)&=&-\frac{2C_F}{N_C}\frac{\pi\alpha\alpha_s e_q^2}{s}\left(
\frac{v}{1-v}+\frac{1-v}{v}\right) ;\nonumber \\
\frac{d\Delta\hat{\sigma}}{dv}(qg\rightarrow \gamma
q)&=&\frac{\pi\alpha\alpha_s e_q^2}{N_C s}\left(
\frac{1-(1-v)^2}{1-v}\right).
\label{eq:afifteen}
\end{eqnarray}
The physical cross section is given by 
\begin{equation}
\Delta\sigma_{frag/c}=\sum_{abd}\int
\Delta f^A_a(x_1,M^2)\Delta f^B_b(x_2,M^2)D_{c/d}(z,M''^2)
\frac{d\Delta\hat{\sigma}}{dv}
(ab\rightarrow \gamma d)dx_1 dx_2 dz dv.
\label{eq:asixteen}
\end{equation}

\subsection{Three-body contributions}

The non-collinear three-body final-state contributions are calculated from 
the expression
\begin{equation}
\Delta\sigma_{3body}=\sum_{abd}\int\Delta f^A_a(x_1,M^2)
\Delta f^B_b(x_2,M^2)d\Delta \hat{\sigma}(ab\rightarrow\gamma cd)dx_1dx_2d\Omega,
\label{eq:aseventeen}
\end{equation}  
with $\Omega$ representing the angles and other variables that are integrated 
over. Whenever an invariant $s_{ij}$ or $t_{ij}$ falls into a collinear or 
soft region of phase space, that contribution from the subprocess is excluded.
The three-body contribution shows no dependence on
the factorization scale of the final-state charm or photon legs,
although we have factored collinear singularities at scales $M''$ and
$M'$, respectively, on these legs of the three-body subprocesses. 
However, Eq.~(\ref{eq:aseventeen}) does contain implicit logarithmic 
dependence on the soft and, in particular, the collinear cutoffs discussed 
in Sec.~II.  Both collinear cutoff and factorization scale
dependences are contained in the pseudo-two-body contributions
discussed above.

\pagebreak

\noindent
\begin{center}
{\large FIGURE CAPTIONS}
\end{center}
\newcounter{num}
\begin{list}%
{[\arabic{num}]}{\usecounter{num}
    \setlength{\rightmargin}{\leftmargin}}

\item (a) Lowest order Feynman diagrams for $\gamma$ plus $c$ quark 
production; $k_1$ and $k_2$ are the four-vector momenta of the photon and
charm quark. (b) Examples of virtual corrections to the lowest order diagrams.
(c) Examples of next-to-leading order three-body final-state diagrams for 
the $g c$ initial state. 

\item (a) Ratio of the polarized to the unpolarized gluon distribution 
$\Delta G(x,\mu)/G(x,\mu$) at ${\mu}^2=100$ GeV$^2$ for the GSA, GSB and GSC 
spin-dependent parton distributions and the CTEQ4M spin-averaged 
distributions. Evolution begins at $\mu_o=$~1.5~GeV, and the number 
of flavors $n_f$ is set to 4 in the evolution.
(b) Same as (a) but for the charm distributions. The charm quark density is
generated from bremsstrahlung of gluons into charm-anticharm pairs.

\item (a) Spin-averaged cross section 
$d\sigma/dp^\gamma_{T}$ as a function of the transverse
momentum of the photon for $p + p \rightarrow \gamma + c + X$ at
$\sqrt{S}=200$ GeV.  The transverse momentum of the charm quark is restricted 
to $p^c_T > 5$ GeV, and the rapidities of the photon and
charm quark are restricted to the interval $-1.0 < y < 1.0$.  Two curves are
drawn.  The solid curve shows the cross section that results from the inclusion 
of all subprocesses through $O(\alpha ^2_s)$, and the dashed curve shows the 
contribution from the dominant $cg$ subprocess, only, through $O(\alpha ^2_s)$.
(b) For the same selections as in (a), the two-spin longitudinal asymmetry 
is shown for the three choices of the spin-dependent gluon density.  In the 
case of GSA, the dot-dashed curve shows the contribution to $A_{LL}$ from 
the $cg$ subprocess only.  
(c) For the same selections as in (a), the two-spin longitudinal asymmetry 
$A_{LL}$ is shown for the GSC spin-dependent gluon density (dotted line), 
along with the contribution from the $cg$ subprocess only.  

\item (a) Spin-averaged cross section $d\sigma/dy^c$ as a function of the 
rapidity of the charm quark for $p + p \rightarrow \gamma + c + X$ at 
$\sqrt{S}=200$ GeV.  The photon rapidity is restricted to 
$-1.0 < y^\gamma < 1.0$, and $4 < p^\gamma_T < 50$ GeV.  The transverse 
momentum of the charm quark is selected to be in the region $p^c_T > 5$ GeV.
The solid curve shows the cross section that results from the inclusion 
of all subprocesses through $O(\alpha ^2_s)$, and the dashed curve shows the 
contribution from the dominant $cg$ subprocess through $O(\alpha ^2_s)$.
(b) For the same selections as in (a), the two-spin longitudinal asymmetry 
is shown for the three choices of the spin-dependent gluon density.  The 
solid, dashed, and dotted curves represent the results from calculations 
based on our modified GSA, GSB, and GSC densities, respectively.  In the GSA  
case, the dot-dashed curve shows the contribution to $A_{LL}$ from 
the $cg$ subprocess only.  

\item (a) Spin-averaged cross section 
$d\sigma/d\Delta y$ as a function of the difference 
$\Delta y = y^c - y^\gamma$ of the rapidities of the photon and charm quark,
for $p + p \rightarrow \gamma + c + X$ at $\sqrt{S}=200$ GeV.  The transverse 
momentum of the photon is selected to be in the interval, 
$4 < p^\gamma_T < 50$ GeV, and its rapidity is restricted to 
$-1.0 < y^\gamma < 1.0$.  The transverse momentum of the charm quark satisfies 
$p^c_T > 5$ GeV.  The solid curve shows the cross section that results from the 
inclusion of all subprocesses, and the dashed curve shows the contribution 
from the dominant $cg$ subprocess only.
(b) For the same selections as in (a), the two-spin longitudinal asymmetry 
is shown for the three choices of the spin-dependent gluon density.  The 
solid, dashed, and dotted curves represent the results from calculations 
based on our modified GSA, GSB, and GSC densities, respectively.  

\end{list}
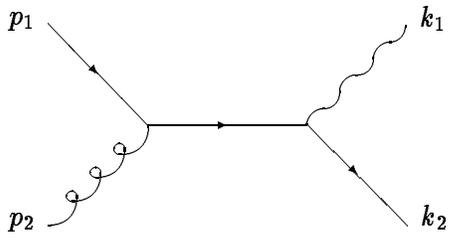
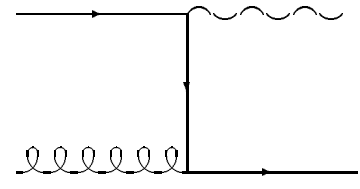
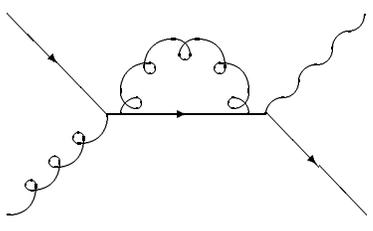
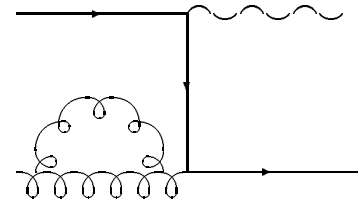
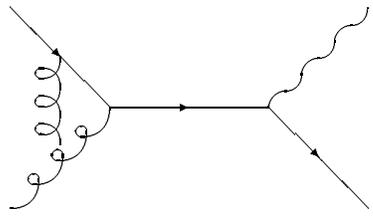
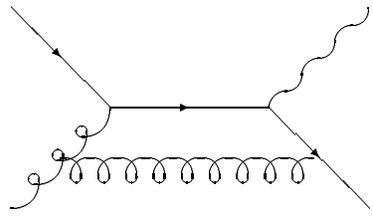
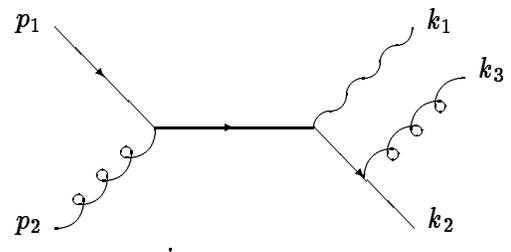
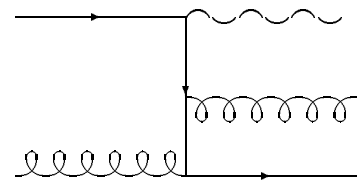
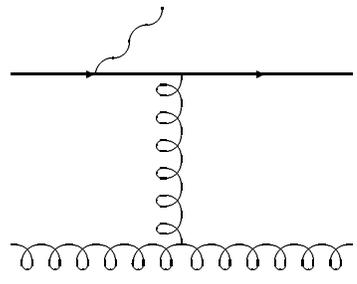
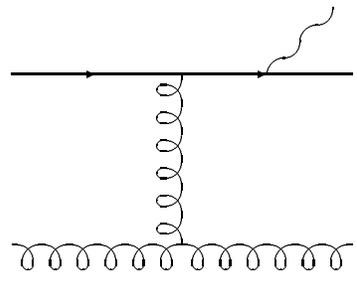
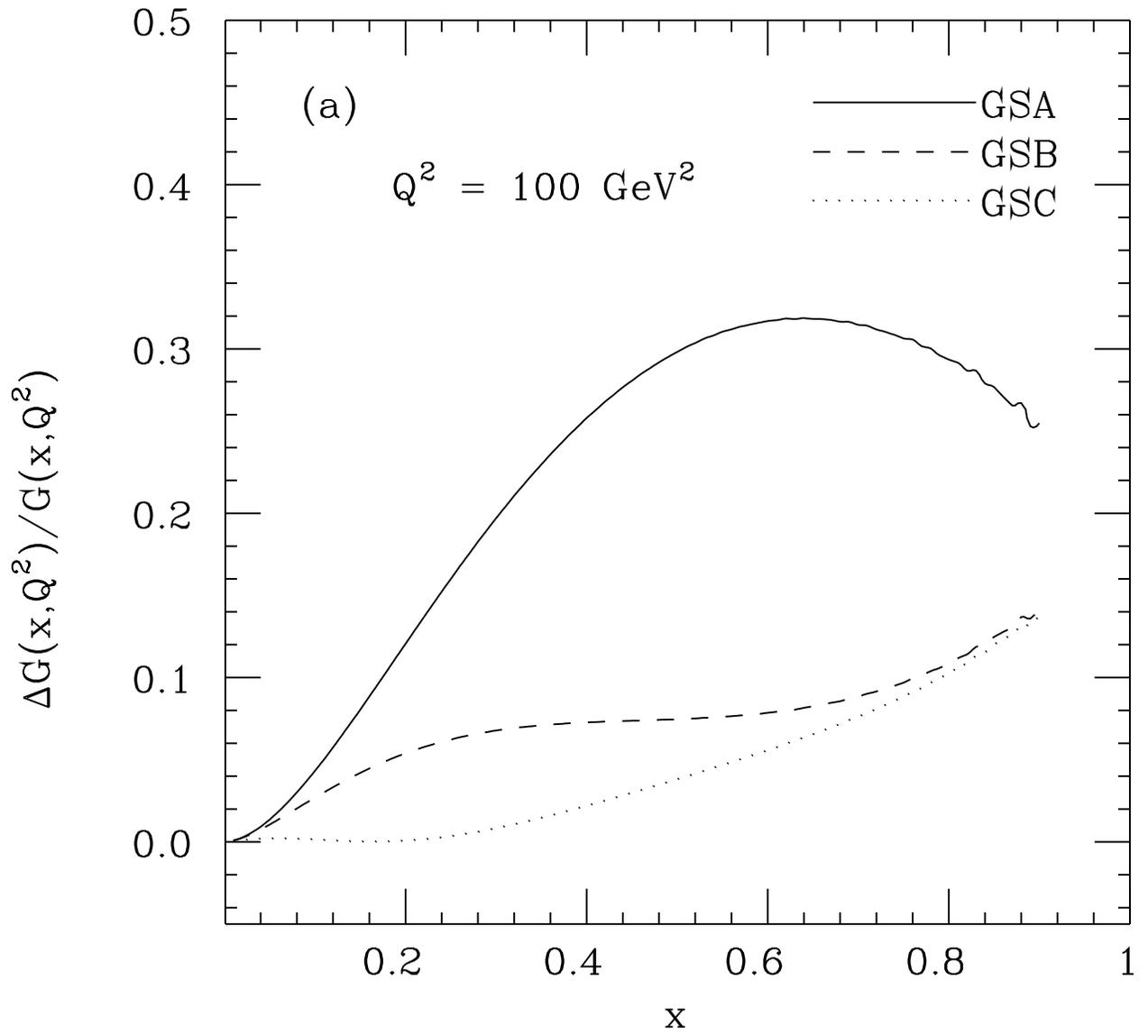
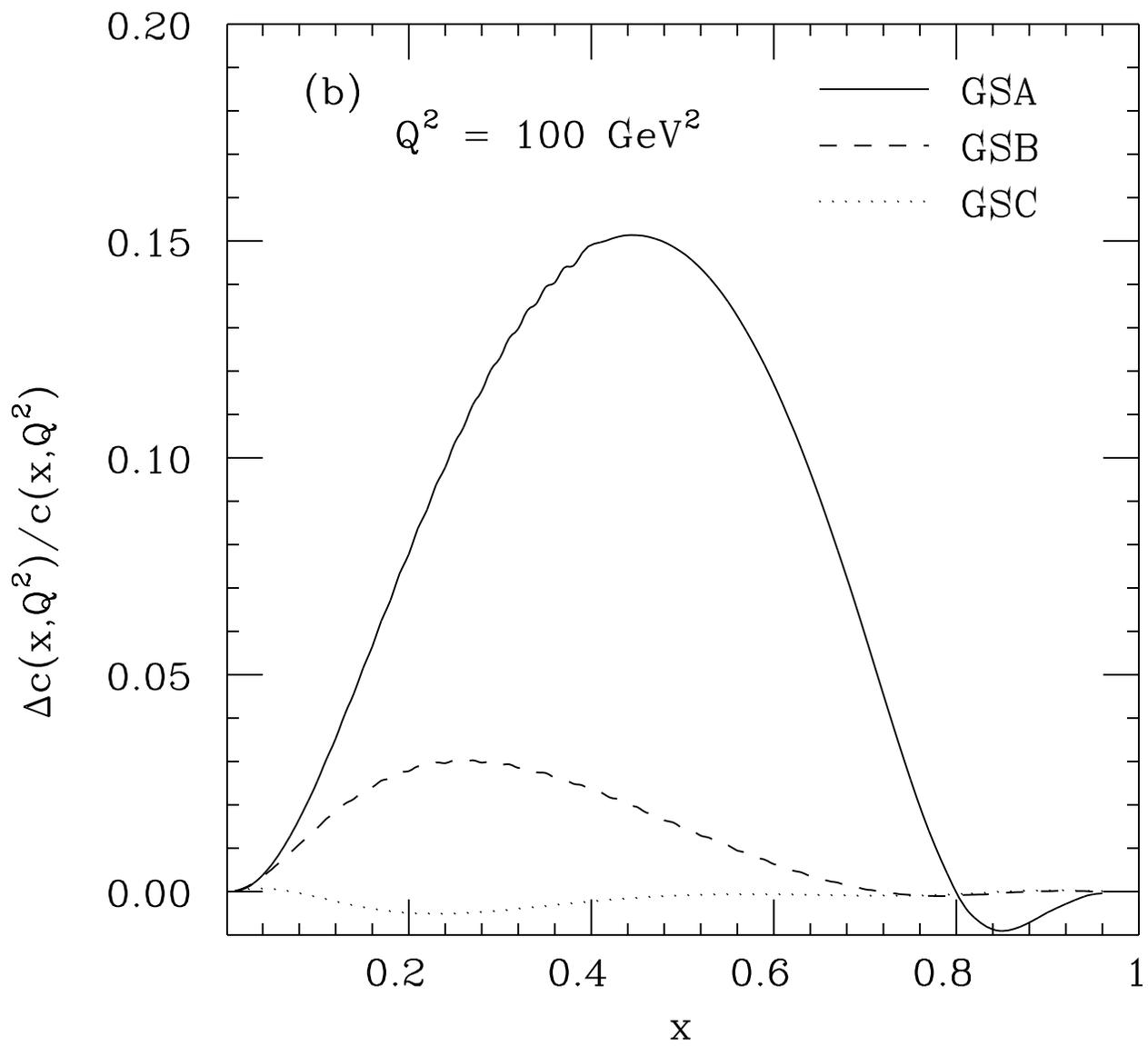
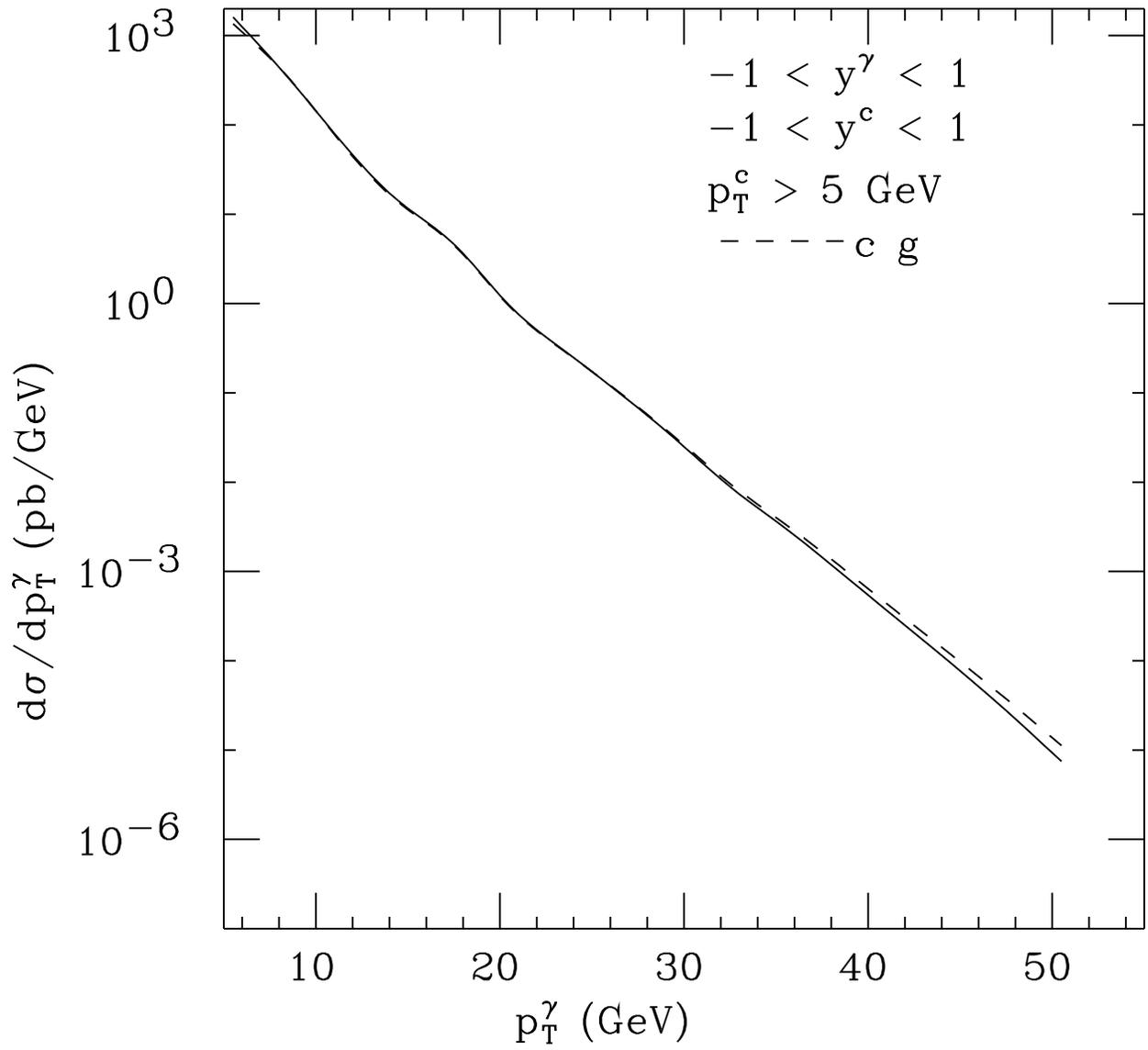
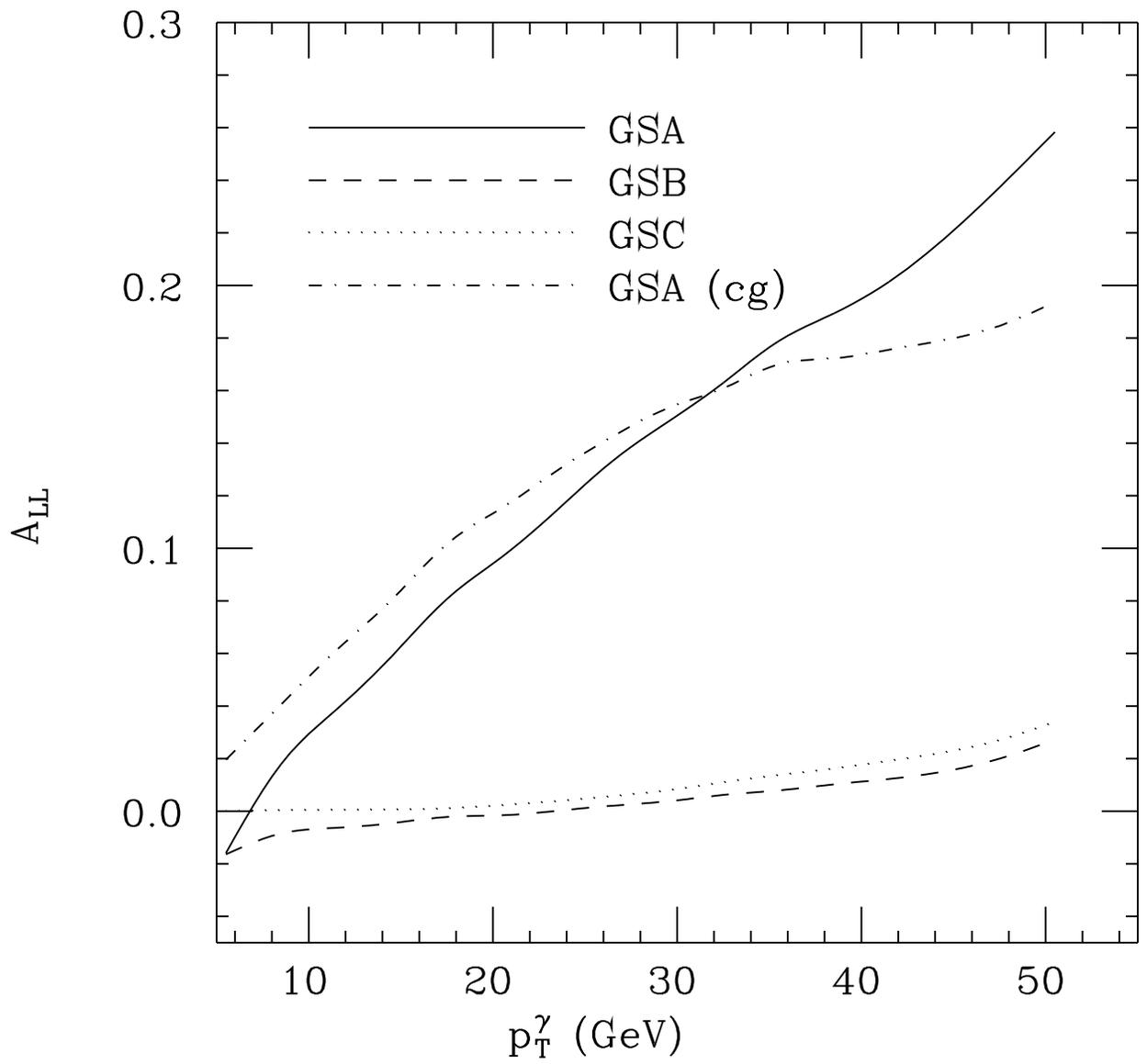
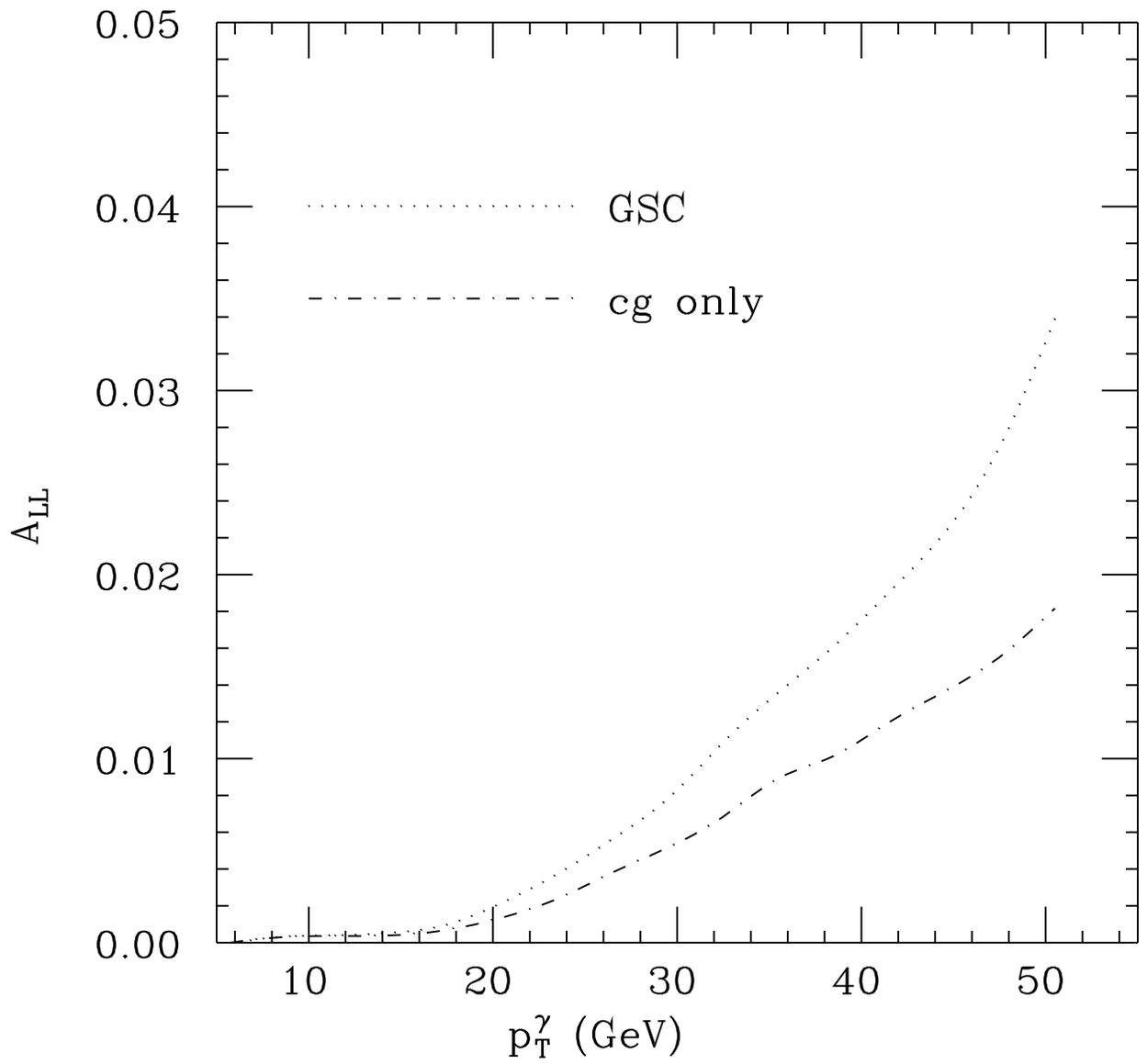
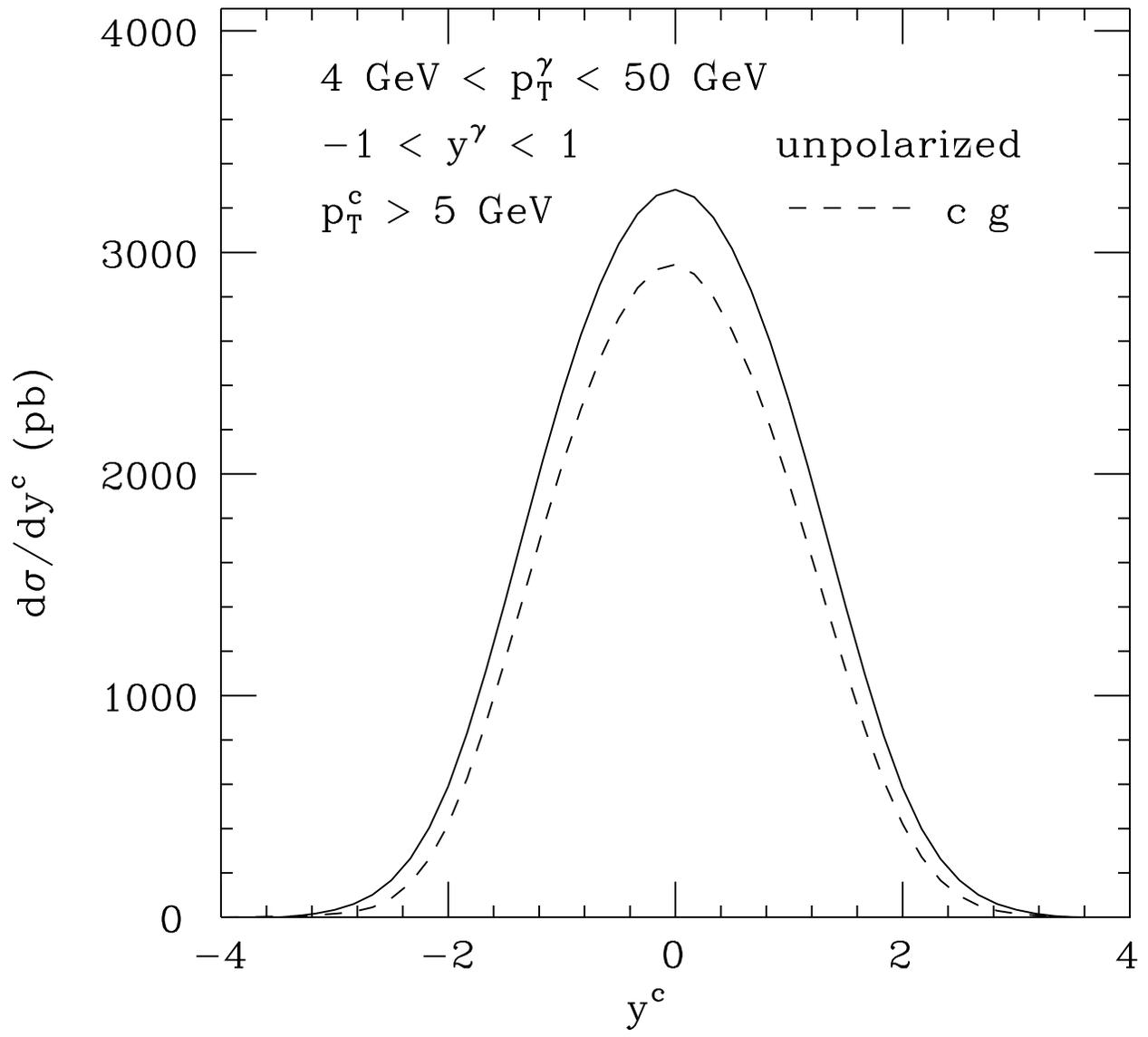
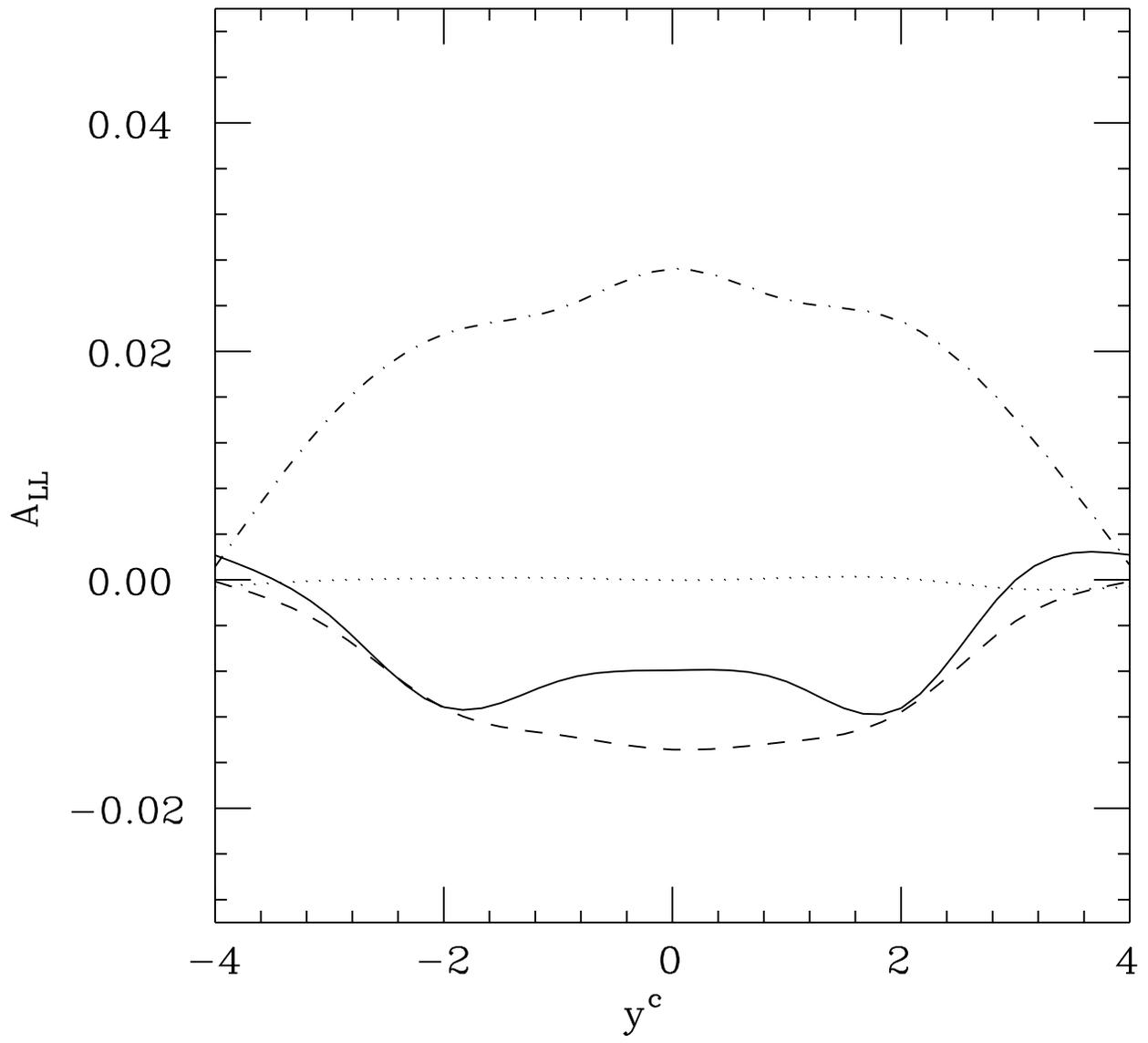
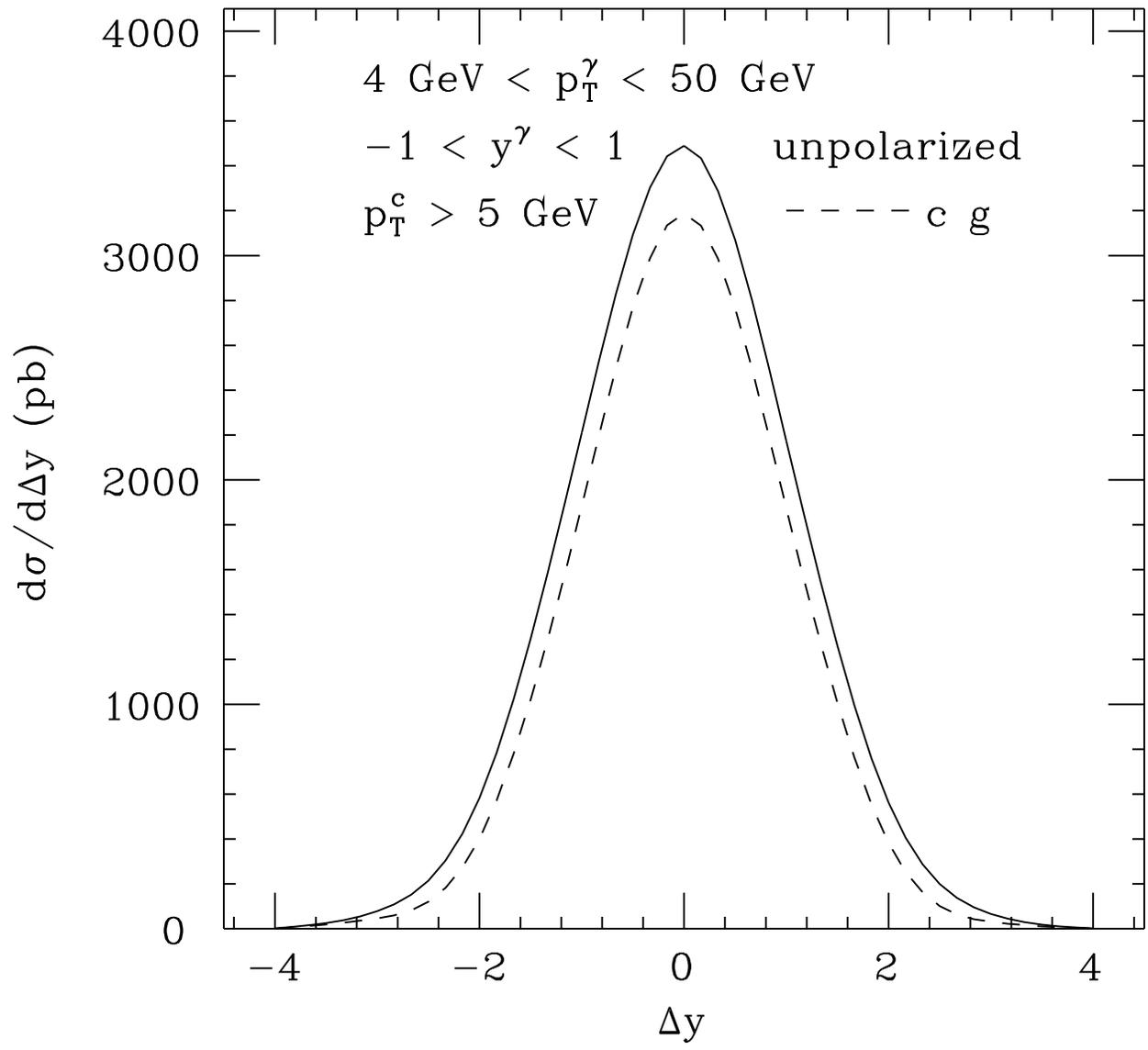
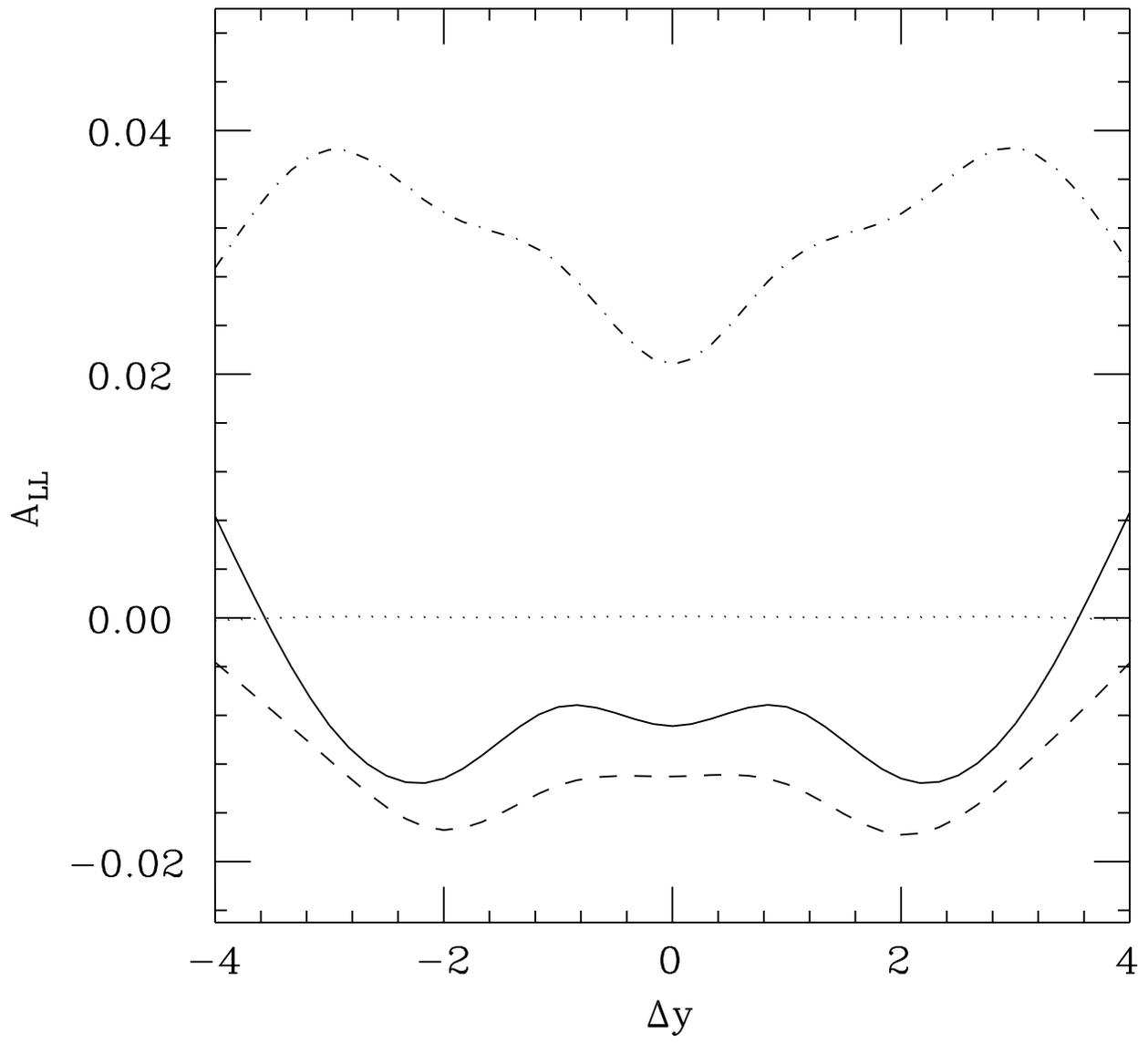
\end{document}